\documentclass[prl,aps,preprint]{revtex4}
\usepackage{myfigcaption2,mycite,mybiblabel,graphicx,overpic}

\def\@biblabel#1{{#1.}} 
\def\@BIBLABEL#1{$^{#1}\m@th$} %

\begin{document}

\title{Experimental realization of Shor's quantum factoring algorithm using
nuclear magnetic resonance}

\vspace*{1ex} \author{Lieven M.K. Vandersypen$^{\dagger,*}$, Matthias
Steffen$^{*,\dagger}$, Gregory Breyta$^\dagger$, Costantino
S. Yannoni$^\dagger$, Mark H. Sherwood$^\dagger$ and Isaac
L. Chuang$^{*,\dagger}$}

\address{\vspace*{1.2ex} {$^\dagger$ IBM Almaden Research Center, San Jose,
CA 95120}\\ {$^*$ Solid State and Photonics Laboratory, Stanford University,
Stanford, CA 94305-4075}\\ }

\maketitle

\def\be{\begin{equation}}
\def\ee{\end{equation}}
\newcommand{\ket}[1]{\mbox{$|#1\rangle$}}
\newcommand{\bra}[1]{\mbox{$\langle#1|$}}
\newcommand{\ang}{\mbox{{\AA}\normalsize}}                              

\small

\vspace*{-2ex}

\noindent {\bf The number of steps any classical computer requires in
order to find the prime factors of an $l$-digit integer $N$ increases
exponentially with $l$, at least using algorithms~\cite{Knuth98a} known 
at present. Factoring large integers is therefore
conjectured to be intractable classically, an observation underlying the
security of widely used cryptographic
codes~\cite{Knuth98a,Koblitz94a}. Quantum computers~\cite{Bennett00a},
however, could factor integers in only polynomial
time, using Shor's quantum factoring
algorithm~\cite{Shor94a,Shor97a,Ekert96a}. Although important for the study of
quantum computers~\cite{Beckman96a}, experimental demonstration of this
algorithm has proved elusive~\cite{Jones01a,Vandersypen00b,Knill00a}. Here we 
report an implementation of the simplest instance of Shor's algorithm:
factorization of ${N\mathbf=\mathbf15}$ (whose prime factors are 3 and 5).
We use seven spin-1/2 nuclei in a molecule as
quantum bits~\cite{Gershenfeld97a,Cory97a}, which can be manipulated with  
room temperature liquid state nuclear magnetic 
resonance techniques. This method of using
nuclei to store quantum information is in principle scalable to many
quantum bit systems~\cite{Vazirani99}, but such scalability is not 
implied by the present work. The significance of our work lies in
the demonstration
of experimental and theoretical techniques for precise control
and modelling of complex quantum computers. In particular, we present a
simple, parameter-free but predictive model of decoherence
effects~\cite{Chuang95a} in our system.}

Shor's factoring algorithm works by utilizing a quantum computer to
quickly determine the period of the function $f(x) = a^x\,
\mbox{mod}\, N$ (the remainder of $a^x$ divided by $N$), where $a$ is
a randomly chosen small number with no factors in common with $N$;
from this period, number-theoretic techniques can be employed to
factor $N$ with high probability~\cite{Shor94a}. The two main
components of the algorithm, modular exponentiation (computation of
$a^x \, \mbox{mod}\, N$) and the inverse quantum Fourier transform
(QFT) take only ${\cal O}(l^3)$
operations~\cite{Shor94a,Shor97a,Ekert96a}. Classically, in contrast,
prime factorization takes ${\cal O}(2^{l^{1/3}})$
operations~\cite{Knuth98a}, which quickly becomes intractable as $l$
increases.

The simplest meaningful instance of Shor's algorithm is factorization
of $N=15$~\cite{Beckman96a} --- the algorithm fails for $N$ even or 
a prime power.  Even
for such a small $N$, quantum factorization poses at present a
significant experimental challenge: it requires coherent control over
seven quantum bits (qubits) in the course of a long sequence of
controlled interactions, even after maximal reduction of the quantum
circuit; including the state initialization, interactions between
almost all pairs of qubits are needed. In comparison with earlier
work~\cite{Jones01a,Vandersypen00b,Knill00a}, this experiment thus
puts extraordinarily high demands on the spin-spin coupling network,
the degree of control over the Hamiltonian and the spin coherence
times. Furthermore, numerically predicting the outcome of these
experiments has been considered impractical due to the enormous size
of the state space transformations, which are described by $\sim 4^7
\times 4^7$ real parameters if decoherence effects are included.

Implementation of the algorithm can be broken into four distinct steps
(Fig.~\ref{fig:circuit}a), with the most complex being the computation
of $f(x) = a^x\, \mbox{mod} \, N$ for $2^n$ values of $x$ in parallel.
Following standard classical circuit techniques, this is performed by
utilizing the identity $a^x = a^{2^{n-1} x_{n-1}} \ldots a^{2 x_1}
a^{x_0}$, where $x_k$ are the binary digits of $x$. Modular
exponentiation thus consists of serial multiplication by $a^{2^{k}}\,
\mbox{mod}\, N$ for all $k$ $(0 \le k \le n-1)$ for which $\ket{x_k} =
\ket{1}$.  The powers $a^{2^{k}}$ can be efficiently pre-computed on a
classical machine by repeated squaring of $a$.  For $N=15$, $a$ may be
$2,4,7,8,11,13$ or $14$. If we happen to pick $a=2,7,8$ or $13$, we
find that $a^4 \,\mbox{mod}\, 15 = 1$, and therefore all $a^{2^{k}}\,
\mbox{mod}\, N = 1$ for $k \ge 2$.  In this case, $f(x)$ simplifies to
multiplications controlled by just two bits, $x_0$ and $x_1$.  If
$a=4,11$ or $14$, then $a^2\, \mbox{mod} \, 15 = 1$, so only $x_0$ is
relevant.  Thus, the first register can be as small as two qubits
($n=2$); however, three qubits ($n=3$) allow for the possibility of 
detecting more periods and thus constitutes a more stringent test of the
modular exponentiation and QFT~\cite{Steffen01a}. Together
with the $m=\lceil \log_2 15 \rceil = 4$
qubits to hold $f(x)$, we need seven qubits in total
(Fig.~\ref{fig:circuit}b).  We implemented this algorithm and tested
it on two representative parameter choices: $a=11$ (an ``easy'' case)
and $a=7$ (a ``difficult'' case).

The custom-synthesized molecule used as the quantum computer for this
experiment contains five $^{19}$F and two $^{13}$C spin-1/2 nuclei as qubits
(Fig.~\ref{fig:molecule}).  In a static magnetic field, each spin has two
discrete energy eigenstates, $\ket{0}$ (spin-up) and $\ket{1}$ (spin-down),
described by the Hamiltonian ${\cal H}_0 = - \sum_i \hbar \omega_i I_{zi}$,
where $\omega_i/2\pi$ is the transition frequency between $\ket{0}$ and $\ket{1}$
and $I_z$ is the $\hat{z}$ component of the spin angular momentum operator.
{\em All} seven spins in this molecule are remarkably well separated in
frequency $\omega_i/2\pi$, and interact pairwise via the $J$-coupling, 
described by ${\cal H}_J = \sum_{i<j} 2 \pi \hbar J_{ij} I_{zi}
I_{zj}$~\cite{Freeman97a}. 

The desired initial state of the seven qubits is
$\ket{\psi_1}=\ket{0000001}$ (Fig.~\ref{fig:circuit}). However,
experimentally we start from thermal equilibrium. The density
matrix is then given by $\rho_{th} = e^{-{\cal H}_0/k_BT}/2^7$, with
$k_BT \gg \hbar \omega_i$ at room temperature so each spin is in a
statistical mixture of $\ket{0}$ and $\ket{1}$
(Fig.~\ref{fig:spectra}a). We converted $\rho_{th}$ into a 7-spin
effective pure state~\cite{Gershenfeld97a,Cory97a} $\rho_1$ via
temporal averaging~\cite{Vandersypen00b} (step 0); $\rho_1$
constitutes a suitable initial state for Shor's factoring algorithm
since it generates the same signal as $|\psi_1\rangle$, up to a
proportionality constant~\cite{Gershenfeld97a,Cory97a}.  While
$\rho_1$ is highly mixed and in fact remains separable under unitary
transforms, the observed dynamics under multiple qubit operations such as
in Shor's algorithm apparently remain hard to simulate
classically~\cite{Braunstein99a,Schack99a,Linden01a}.

The quantum circuit of Fig.~\ref{fig:circuit} was realized with a
sequence of $\sim 300$ ($a=7$) spin-selective radio-frequency (RF)
pulses separated by time intervals of free evolution under the
Hamiltonian (Fig.~\ref{fig:sequence}). The pulse sequence is designed
such that the resulting transformations of the spin states correspond
to the computational steps in the algorithm. Upon completion of this
sequence, we estimate the state of the first three qubits, $\rho \sim
\sum_c w_c \ket{c2^3/r}\bra{c2^3/r}$, via nuclear magnetic resonance
(NMR) spectroscopy.  In the experiment, an ensemble of independent
quantum computers rather than a single quantum computer was used, so
the measurement gives the bitwise average value of $8c/r$, instead of
a sample of $8c/r$.  This is sufficient to determine $r$ in the
present experiment, but for larger $N$ a continued fractions algorithm
will need to be performed on the quantum
computer~\cite{Gershenfeld97a} requiring additional qubits. From 
$r$, at least one factor of $N$
is given by the greatest common denominator (gcd) of $a^{r/2} \pm 1$
and $N$ (with probability greater than $1/2$); the gcd can be computed
efficiently using Euclid's algorithm on a classical
computer~\cite{Koblitz94a} .

The experimental spectra acquired upon completion of the easy case
($a=11$) of Shor's algorithm (Fig.~\ref{fig:spectra}c) clearly
indicate that qubits 1 and 2 are in $\ket{0}$ (spectral lines up), and
that 3 is in an equal mixture of $\ket{0}$ and $\ket{1}$ (lines up and
down, and the integral of the spectrum equal to zero).  With qubit 3
the most significant qubit after the inverse
QFT~\cite{Coppersmith95a}, the first register is thus in a mixture of
$\ket{000}$ and $\ket{100}$, or $\ket{0}$ and $\ket{4}$ in decimal
notation. The periodicity in the amplitude of $\ket{y}$ is thus 4, so
$r = 2^n / 4 = 2$ and we find that $\mbox{gcd} (11^{2/2} \pm 1, 15) =
3,5$. The prime factors thus unambiguously derive from the output
spectra.

From analogous spectra for the difficult case ($a=7$;
Fig.~\ref{fig:spectra}d), we see that qubit 1 is in $\ket{0}$, and
qubits 2 and 3 are in a mixture of $\ket{0}$ and $\ket{1}$. The
register is thus in a mixture of $\ket{000}$, $\ket{010}$, $\ket{100}$
and $\ket{110}$, or $\ket{0}$, $\ket{2}$, $\ket{4}$ and $\ket{6}$. The
periodicity in the amplitude of $\ket{y}$ is now 2, so $r = 8/2 = 4$
and $\mbox{gcd} (7^{4/2} \pm 1, 15) = 3,5$. Thus, even after the long
and complex pulse sequence of the difficult case, the experimental
data conclusively indicate the successful execution of Shor's
algorithm to factor $15$.

Nevertheless, there are obvious discrepancies between the measured and
ideal spectra, most notably for the difficult case. Using a numerical
model, we have investigated whether these deviations could be caused
by decoherence.  A full description of relaxation for the seven
coupled spins involves almost $4^7\times4^7$ degrees of freedom and requires
knowledge of physical properties of the molecule which are not
available~\cite{Vold78a,Jeener82a}. In order to get a first estimate
of the impact of decoherence during the factoring pulse sequence we 
assume that each spin experiences 
independent stochastic relaxation with correlation time scales 
$\ll 1/\omega_i$. This
permits the use of the phenomenological Bloch equations~\cite{Bloch46a},
with just two time constants per spin ($T_1$ and $T_2$).  We
implemented this decoherence model for seven coupled spins via the
operator sum representation~\cite{Kraus83a} $\rho \mapsto \sum_k E_k
\rho E_k^\dagger$ ($\sum_k E_k^\dagger E_k = I$), starting from
existing single spin models~\cite{Nielsen00b} of generalized amplitude
damping (GAD, $T_1$),
\begin{eqnarray}
\nonumber
\hspace*{-1cm}
E_0 = \sqrt{p} \left[\matrix{1 & 0 \cr 0 & \sqrt{1-\gamma} }\right] 
&\,,& \;\;
E_1 = \sqrt{p} \left[\matrix{0 & \sqrt{\gamma} \cr 0 & 0 }\right] \,,\\
E_2 = \sqrt{1-p} \left[\matrix{\sqrt{1-\gamma} & 0 \cr 0 & 1}\right] 
&\,,& \;\;
E_3 = \sqrt{1-p} \left[\matrix{0 & 0 \cr \sqrt{\gamma} & 0 }\right] \,.
\label{eq:opsumrep_gen_ad}
\end{eqnarray}
and phase damping (PD, related to $T_2$),
\begin{eqnarray}
E_0 = \sqrt{\lambda} \left[\matrix{1 & 0 \cr 0 & 1}\right] \,, \;\;
E_1 = \sqrt{1-\lambda} \left[\matrix{1 & 0 \cr 0 & -1}\right] \,,
\label{eq:opsumrep_pd}
\end{eqnarray} 
with $\gamma = 1-e^{-t/T_1}$, $p = 1/2 + \hbar\omega/4k_BT$ and
$\lambda \sim (1 + e^{-t/T_2})/2$.
The following observations simplify the extension of these separate
single spin descriptions to an integrated model for seven spins: (1)
GAD (and PD) error operators acting on different spins commute; (2)
the $E_k$ for GAD commute with the $E_k$ for PD when applied to
arbitrary $\rho$; and (3) PD commutes with the ideal unitary time
evolution $e^{-i{\cal H}t}$ (${\cal H} = {\cal H}_0 + {\cal H}_J$).
However, GAD does {\em not} commute with $e^{-i{\cal H}t}$, and PD and
GAD do {\em not} commute with the ideal unitary evolution during RF
pulses.  Nevertheless, we have treated these as commuting
transformations, such that all of the processes which occur
simultaneously can be modeled sequentially.

Specifically, the model simulates a delay time of duration $t_d$ by $e^{- i
{\cal H} t_d}$ followed by GAD acting on spin 1 for a duration $t_d$, then
GAD acting on spin 2 and so forth, followed by PD acting serially on each
spin. Similarly, a shaped pulse of duration $t_p$ was modeled by a delay
time of duration $t_p$ ($e^{-i {\cal H}_0 t_p}$, GAD and PD) followed by an
instantaneous pulse.  Using this simple model, the 7-spin simulation of the
complete Shor pulse sequence, including 36 temporal averaging sequences,
required only a few minutes on an IBM quad {\sc power}3-II processor
machine.  Measured values of $T_1$ and $T_2$ (Fig.~\ref{fig:molecule}) were
used in the model.

The output spectra predicted by this parameter-free decoherence model are
also shown in Figs.~\ref{fig:spectra}c and d. While some discrepancies
between the data and the simulations remain (due to the approximations in
the model as well as due to experimental imperfections such as RF
inhomogeneity, imperfect calibrations, incomplete field drift compensation
and incomplete unwinding of coupled evolution during the RF
pulses~\cite{Steffen01a}), the model agrees well with the
large non-idealities of the data.
The predictive value of the model was further confirmed via independent 
test experiments.

This is the first NMR quantum computation experiment for which decoherence
is the dominant source of errors~\cite{Jones01a}; the demands of Shor's
algorithm clearly push the limits of the current molecule, despite its
exceptional properties.  At the same time, the good agreement between the
measured and simulated spectra suggests that the degree of unitary control
in the experiment was very high, which bodes well for related proposed
implementations of quantum computers~\cite{Kane98a,Loss98a}. Finally, our
parameter-free decoherence model, a predictive tool for modeling quantum
errors in this complex system, provides an avenue for future design
simulation of quantum computers.

\vspace*{2ex}
\noindent
{\sf Methods}\\
\setlength\parskip{-.3ex}
\vspace*{-2ex}

Experiments were performed at the IBM Almaden Research Center with an
11.7 T ($500$ MHz) Oxford Instruments magnet, a custom-modified
four-channel Varian Unity INOVA spectrometer, and a Nalorac HFX probe.
We extended the techniques of Ref.~\incite{Vandersypen00b} for serving
multiple nuclei per channel, for reducing cross-talk between RF pulses
on different spins and for sending simultaneous pulses.  We used
spin-selective Hermite-180 and Gaussian-90 pulses~\cite{Freeman97a},
shaped in 256 steps, with a duration of $0.22$ to $\sim 2$ ms. A
technique to compensate for coupling effects during the selective
pulses was developed and implemented via ``negative delay'' times
before {\em and} after the pulse. The amount of negative evolution
needed depends on the pulse shape and was pre-computed via numerical
simulations.

To create an effective pure ground state of {\em all seven} spins,
which has never been done before, we used a two-stage extension of the
scheme of Ref.~\incite{Vandersypen00b}, necessary because
$\omega_{^{13}C}$ is very different from $\omega_{^{19}F}$. The five
$^{19}$F spins are made effective pure via the summation of nine
experiments, each with a different sequence of $CN_{ij}$ and $N_i$
gates ($CN_{ij}$ stands for a controlled-{\sc not} operation, which
flips the target qubit $j$ if and only if the control $i$ is in
$\ket{1}$; $N_i$ simply flips $i$)~\cite{Nielsen00b}. These nine
experiments are executed four times, each time with different
additional $CN_{ij}$ and $N_i$, such that the two $^{13}$C spins are
also made effective pure. Summation of the $4 \times 9 = 36$
experiments along with a {\sc not} on spin 7 gives $\rho_1$. 
The state preparation sequences were
designed to be as short as possible ($\sim 200$ ms) by making optimal
use of the available coupling network~\cite{Steffen01a}.

Multiplication of $y=1$ by $a\,\mbox{mod}\,15$ controlled by $x_0$
(qubit 3) was replaced by controlled-addition of
$(a-1)~\mbox{mod}~15$. For $a=11$, this is done by $CN_{34} CN_{36}$
and for $a=7$ by $CN_{35} CN_{36}$ (gates $A$ and $B$ of
Fig.~\ref{fig:circuit}b). Multiplication of $y$ by $7^2~\mbox{mod}~15$
is equivalent to multiplication of $y$ by $4~\mbox{mod}~15$, which
reduces to swapping $y_0$ with $y_2$ and $y_1$ with $y_3$. Both {\sc
swap} operations must be controlled by $x_1$, which can be
achieved~\cite{Vandersypen00a} via gates $C,D,E$ and $F,G,H$ of
Fig.~\ref{fig:circuit}b. This quantum circuit can be further
simplified by a quantum analogue to peephole compiler
optimization~\cite{Aho86a}, which should become standard in future
quantum compilers: (1) the control qubit of gate $C$ is $\ket{0}$, so
$C$ was suppressed; (2) similarly, $F$ was replaced by $N_5$; (3)
gates $H$ and $E$ are inconsequential for the final state of the first
register, so they were omitted; (4) the targets of the doubly
controlled {\sc not} gates $D$ and $G$ are in a basis state, so they
can be implemented as $CY^\dagger_{24} CZ^2_{64} CY_{24}$ and
$CY^\dagger_{27} CZ^2_{57} CY_{27}$ ($CZ_{ij}$ stands for a $90^\circ$
$\hat{z}$ rotation of $j$ if and only if $i$ is in $\ket{1}$); (5) the
refocusing schemes were kept as simple as possible. To this end, $A$
was carried out after $E$. We did refocus inhomogeneous dephasing for
all spins in the transverse plane. Residual couplings with the
cyclopentadienyl protons were decoupled by continuous on-resonance low
power irradiation using a separate power amplifier and additional
power combiners and RF filters. After all these simplifications, the
pulse sequence for $7^x~\mbox{mod}~15$ was $\sim 400$ ms long. The
inverse QFT was implemented as shown in Fig.~\ref{fig:circuit}b and
took $\sim 120$ ms. The duration of the complete sequence for the Shor
algorithm was thus up to $\sim 720$ ms.  A detailed report on these
methods will be published elsewhere~\cite{Steffen01a}.

~\\

{\sf Acknowledgements}\\
\vspace*{-2ex}

We thank X. Zhou and J. Preskill for useful discussions, J. Smolin for
the use of his IBM workstation, D. Miller for help with spectral
analysis, A. Schwartz and his team at Varian for their generous technical
assistance, and J. Harris, W. Risk and H. Coufal for their
support. L.V. gratefully acknowledges a Yansouni Family Stanford
Graduate Fellowship. This work was supported by DARPA under the NMRQC
initiative.

Correspondence and requests for materials should be addressed to ILC
(e-mail: ichuang@media.mit.edu).


\clearpage

\begin{figure}[h]
\begin{center}
\end{center}
\caption{{\bf a.} Outline of the quantum circuit for Shor's algorithm. Wires
represent qubits and boxes represent operations. Time goes from left to
right. {\bf (0)} Initialize a first register of $n = 2 \lceil \log_2 N
\rceil$ qubits to $\ket{0}\otimes\ldots\otimes\ket{0}$ (for short $\ket{0}$)
and a second register of $m = \lceil \log_2 N \rceil$ qubits to
$\ket{0}\otimes\ldots\otimes\ket{0}\otimes\ket{1}$ ($\ket{1}$). {\bf (1)}
Apply a Hadamard transform $H$ to the first $n$ qubits, so the first
register reaches $\sum_{x=0}^{2^n-1} \ket{x}/ \sqrt{2^n}$. {\bf (2)}
Multiply the second register by $f(x) = a^x\, \mbox{mod} \, N$ 
(for some random $a<N$
which has no common factors with $N$), to get
$
\ket{\psi_2} = 
\sum_{x=0}^{2^n-1} \ket{x}\ket{1 \times a^x \,\mbox{mod}\, N} / \sqrt{2^n}\,.
$
Since the first register is in a superposition of $2^n$ terms $\ket{x}$, the
modular exponentiation is computed for $2^n$ values of $x$ in parallel. {\bf
(3)} Perform the inverse QFT on the first
register~\protect\cite{Coppersmith95a}, giving
$
\ket{\psi_3} 
= 
\sum_{y=0}^{2^n-1} \sum_{x=0}^{2^n-1} e^{2\pi i x y/2^n} \ket{y}
\ket{a^x\, \mbox{mod}\, N} / 2^n
$,
where interference causes only terms $\ket{y}$ with $y = c 2^n / r$ (for
integer $c$) to have a substantial amplitude, with $r$ the period of
$f(x)$. {\bf (4)} Measure the qubits in the first register. On an ideal
single quantum computer, the measurement outcome is $c 2^n/r$ for some $c$
with high probability, and $r$ can be quickly deduced from $c 2^n/r$ on a
classical computer via continued fractions~\protect\cite{Koblitz94a}.
{\bf b.} Detailed quantum circuit for the case $N=15$ and $a=7$. Control
qubits are marked by $\bullet$; $\oplus$ represents a {\sc not} operation
and 90 and 45 represent $\hat{z}$ rotations over these angles. The gates
shown in dotted lines can be removed by optimization and the
gates shown in 
dashed lines can be replaced by simpler gates (see the methods
section).}
\label{fig:circuit}
\end{figure}

\begin{figure}[h]
\begin{center}
\end{center}
\caption{Structure and properties of the quantum computer molecule, a
perfluorobutadienyl iron complex with the inner two carbons
$^{13}$C-labeled. Based on the measured $J_{^{13}\mathrm{C}^{19}\mathrm{F}}$
values, we concluded that the placement of the iron is as shown, different
than derived in Ref.\protect\incite{Green68a} from infrared spectroscopy.
The table gives the $\omega_i/2\pi$ at $11.7$ T, relative to a reference
frequency of $\sim 470$ MHz and $\sim 125$ MHz for $^{19}$F and $^{13}$C 
respectively [Hz], 
the longitudinal ($T_1$, inversion recovery) and
transverse ($T_2$, estimated from a single spin-echo sequence) relaxation
time constants [s], and the $J$-couplings [Hz]. Ethyl
(2-$^{13}$C)bromoacetate (Cambridge Isotope Laboratories, Inc.) was 
converted to ethyl
2-fluoroacetate by heating with AgF followed by hydrolysis to sodium
fluoroacetate using NaOH in MeOH.  This salt was converted to
1,1,1,2-tetrafluoroethane using MoF$_6$ and was subsequently treated with
two equivalents of BuLi followed by I$_2$ to provide trifluoroiodoethene.
Half of the ethene was converted to the zinc salt which was recombined with
the remaining ethene and coupled using Pd(Ph$_3$P)$_4$ to give
(2,3-$^{13}$C)hexafluorobutadiene.
The end product was obtained by reacting
this butadiene with the anion obtained from treating
[($\pi$-C$_5$H$_5$)Fe(CO)$_2$]$_2$ with sodium
amalgam~\protect\cite{Green68a}. The product was purified with column
chromatography, giving a total yield of about $5\%$.
The sample at $0.88 \pm 0.04$ mole $\%$ in diethyl ether-d10 was dried 
using $3$ $\ang$ molecular sieves, filtered through a 0.45 $\mu$m syringe 
filter and flame sealed in the NMR sample tube using three freeze-thaw 
vacuum degassing cycles. All experiments were performed at $30^\circ$C.
}
\label{fig:molecule}
\end{figure}

\begin{figure}[h]
\begin{center}
\end{center}
\caption{Pulse sequence for implementation of the quantum circuit of
Fig.~\ref{fig:circuit} for $a=7$. The tall red lines represent
$90^\circ$ pulses selectively acting on one of the seven qubits (horizontal
lines) about positive $\hat{x}$ (no cross), negative $\hat{x}$ (lower
cross) and positive $\hat{y}$ (top cross). Note how single $90^\circ$ pulses
correspond to Hadamard gates and pairs of such pulses separated by delay times 
correspond to two-qubit gates. The smaller blue lines
denote $180^\circ$ selective pulses used for refocusing~\cite{Leung00a}, 
about positive (darker shade) and negative $\hat{x}$ (lighter shade).
Rotations about $\hat{z}$ are denoted by smaller and thicker green 
rectangles and were implemented with frame-rotations.
Time delays are not drawn to scale.
The vertical dashed black lines visually separate the steps of the
algorithm; step (0) shows one of the 36 temporal averaging sequences.}
\label{fig:sequence}
\end{figure}

\begin{figure}[h]
\caption{NMR spectra at different stages in the computation.
{\bf a.} Experimentally measured thermal equilibrium spectra (real
part), acquired after a read-out pulse on spin $i$ has tipped the spin from
$\ket{0}$ ($+\hat{z}$) or $\ket{1}$ ($-\hat{z}$) into the
$\hat{x}-\hat{y}$-plane, where it induces a voltage oscillating at
$\omega_i/2\pi + \sum_j \pm J_{ij}/2$ (where the sign depends on the state of
the other spins) in a transverse RF coil placed nearby the sample. This
voltage was recorded by a phase-sensitive detector and Fourier transformed
to obtain a spectrum, with the phase set such that positive (negative) lines
correspond to a spin in $\ket{0}$ ($\ket{1}$) before the readout
pulse. Frequencies are in Hertz, and with respect to $\omega_i/2\pi$. {\bf b.}
Experimental spectra for the effective pure ground state. As desired, only one line is
retained in each multiplet with its position depending on strength and sign
of the J-couplings. Here, the transition corresponds to all other spins
in $\ket{0}$. The state $\rho_1$ is obtained from this state by applying
a {\sc not} on spin 7.  {\bf c.} Output spectra of the easy case of 
Shor's algorithm
($a=11$). The top traces are the ideally expected spectra, the middle traces
are the experimental data, and the bottom traces are simulations which
incorporate decoherence effects (vide infra). Each trace was rescaled
separately. {\bf d.} Similar set of spectra as in c, but for the difficult
case ($a=7$).}
\label{fig:spectra}
\end{figure}

\clearpage
\includegraphics*[width=6.5in]{figure1}

\includegraphics*[width=6.5in]{figure2}

\includegraphics*[width=7in]{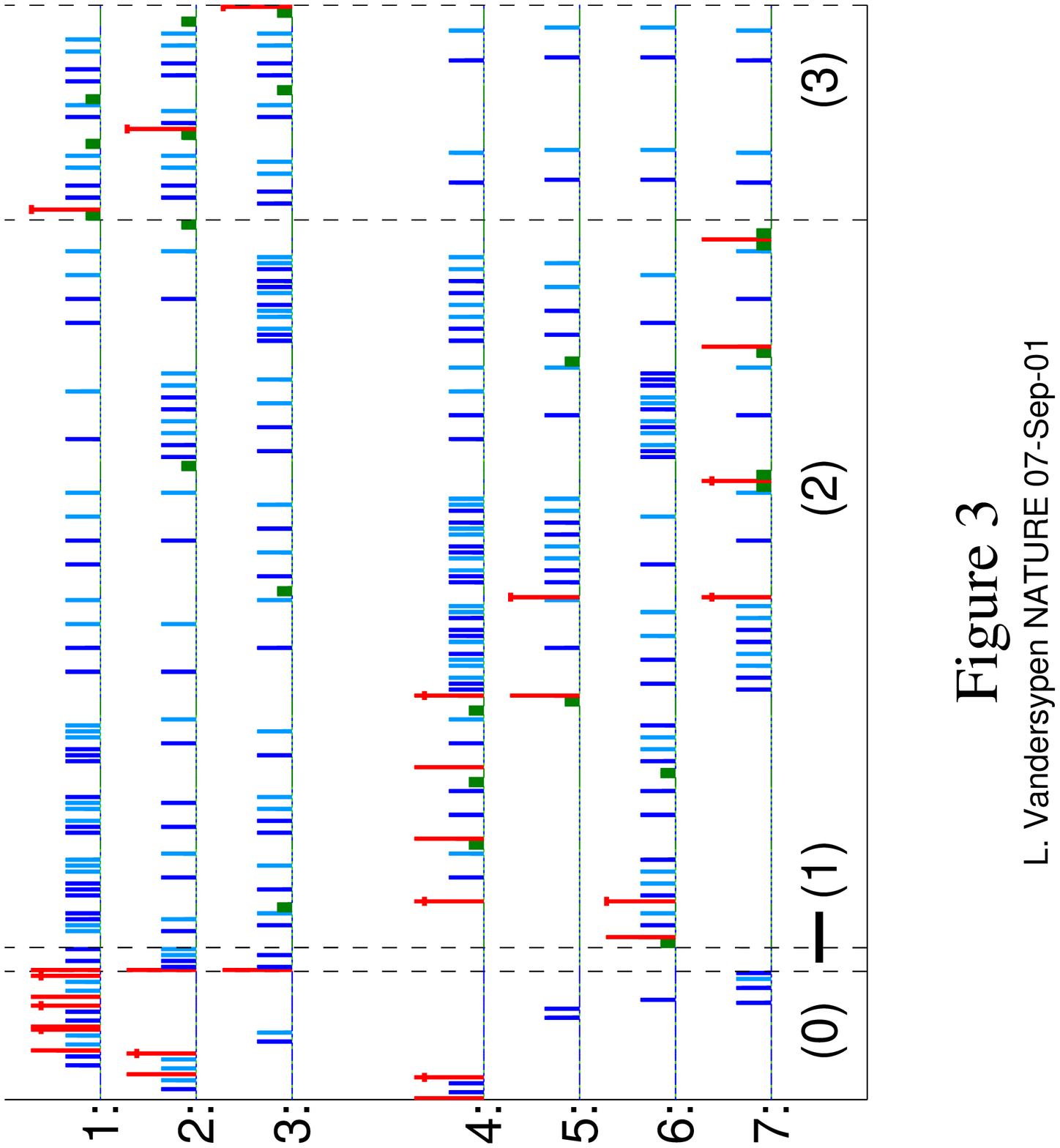}

\centerline{\includegraphics*[width=7in]{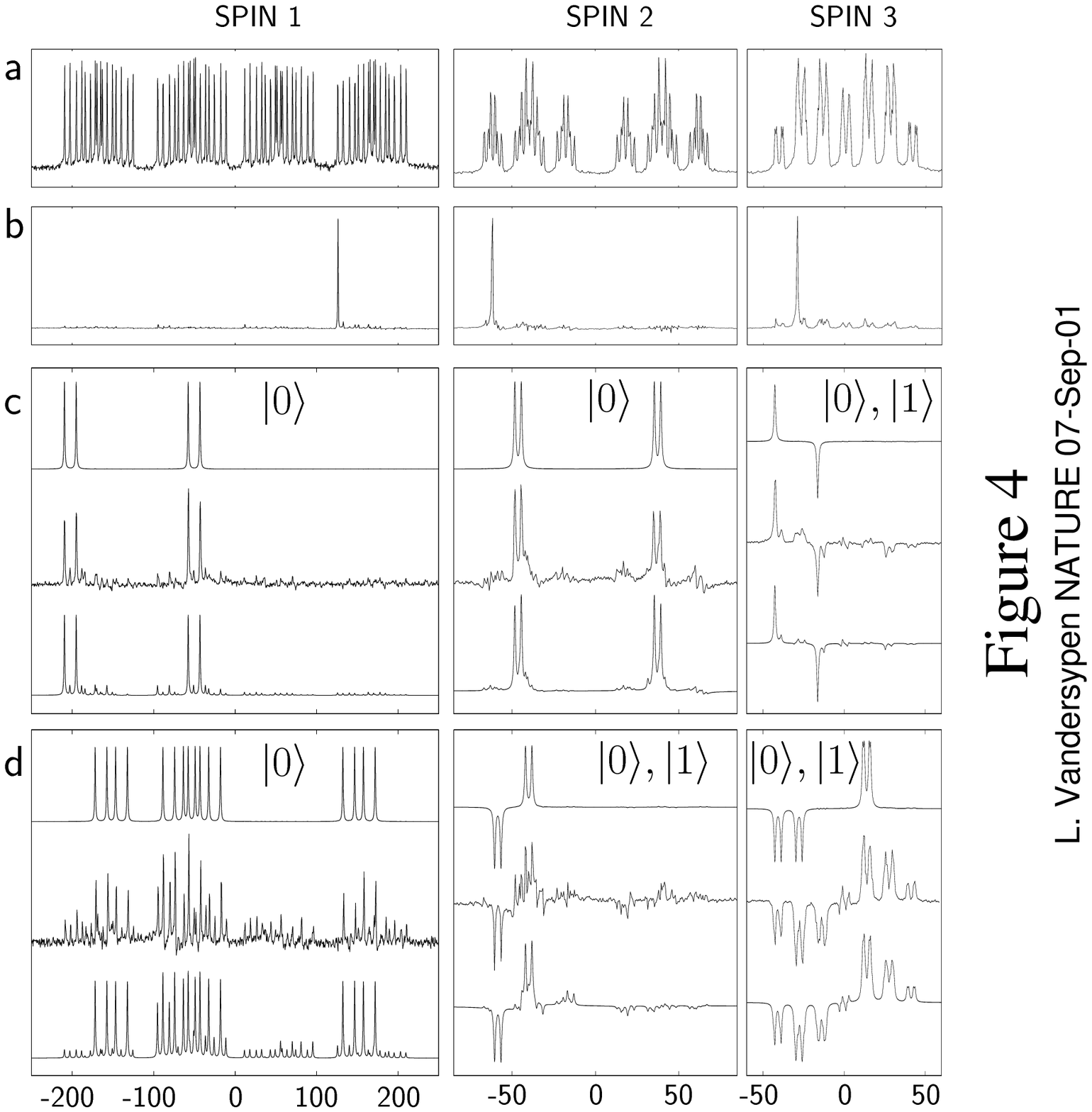}}

\end{document}